\documentclass[
    ,draft            
  ]
  {aipproc}

\layoutstyle{6x9}

\usepackage{graphicx}
\usepackage{bm}
\usepackage{amsmath}
\usepackage{hhline}
\usepackage{amssymb}
\usepackage{times}

\begin{document}

\title[The black--disk limit]{The black--disk limit in high--energy \\
$ep$ and $pp$ scattering\footnote{Notice: This manuscript has been
authored by The Southeastern Universities Research Association,
Inc. under Contract No. DE-AC05-84150 with the U.S. Department of
Energy. The United States Government retains and the publisher, by
accepting the article for publication, acknowledges that the United
States Government retains a non-exclusive, paid-up, irrevocable, world
wide license to publish or reproduce the published form of this
manuscript, or allow others to do so, for United States Government
purposes.}}


\author{M.~Strikman}{
  address={Department of Physics, Pennsylvania State
University, University Park, PA 16802, USA}
}

\author{C.~Weiss}{
  address={Theory Center, Jefferson Lab, Newport News, VA 23606, USA}
}

\begin{abstract} 
We review a recently proposed framework for studying the approach to
the unitarity limit in small--$x$ processes in $ep$ and $pp$
scattering within the DGLAP approximation. Our formulation is based on
the correspondence between the standard QCD factorization theorems for
hard processes and the dipole picture of small--$x$ scattering in the
leading logarithmic approximation.  It allows us to incorporate
information about the transverse spatial distribution of gluons in the
proton (GPD) from exclusive vector meson production at HERA. We show 
that the interaction of small--size color singlet configurations with the
proton approaches the ``black--disk limit'' (BDL) due to the growth of
the DGLAP gluon density at small $x$.  This new dynamical regime is
marginally visible in diffractive DIS at HERA, and will be fully
reached in central $pp$ collisions at LHC.
\end{abstract}

\maketitle


\section{Introduction}

At TeV energies, strong interactions enter a regime in which 
hadron--hadron and electron/photon--hadron cross sections can become 
comparable to the ``geometric size'' of hadrons, and unitarity becomes an 
essential feature of the dynamics. In QCD, the approach to the unitarity 
regime is driven by the increase of the gluon density in hadrons at 
small $x$. Theoretical studies so far have mostly focused on 
incorporating unitarity effects in QCD evolution equations based 
on the large $\log (1/x)$ approximation (BFKL evolution)
\cite{Balitsky:1995ub,Jalilian-Marian:1997jx,Kovchegov:1999yj}.
However, such calculations are of practical relevance only in situations
where the amount of glue produced by $\log (1/x)$--enhanced radiation
exceeds by far that originating from other sources. 

It is known that there is a large non-perturbative gluon density 
in the proton at a low scale ($Q^2 \sim 4 \, \text{GeV}^2$). 
Its physical origin may be seen in the spontaneous breaking 
of chiral symmetry, which implies that a significant fraction 
of the nucleon's momentum should be carried
by gluon fields. This large non-perturbative gluon density is
the reason behind the success of the DGLAP evolution equations ---
based on the large $\log Q^2$ approximation and retaining
only first (LO) or second (NLO) powers of $\log (1/x)$ ---
in describing the HERA DIS data. Theoretical studies indicate
that $\log (1/x)$--resummation effects are small 
down to $x \sim 10^{-5}$ \cite{Salam:2005yp}. At the
same time, DGLAP evolution, starting from the non-perturbative
gluon density at the low scale, 
leads to a rapid growth of the gluon density at small $x$.
This strongly suggests that unitarity effects become important
long before the region of $\log (1/x)$ dominance is reached.

Recently, we have proposed a simple framework for studying the
approach to the unitarity limit within the DGLAP approximation
\cite{Frankfurt:2003td,Frankfurt:2005mc}. Our formulation makes use of
the correspondence between the standard QCD factorization theorems for
hard processes and the dipole picture of small--$x$ scattering in the
leading logarithmic approximation. This framework allows one to
approach the question of unitarity at small $x$ on the basis of the
vast amount of data on hard processes in $ep$ scattering at HERA
(inclusive, diffractive, and exclusive) and $\bar p p$ scattering at
the Tevatron.  In particular, we incorporate information about the
transverse spatial distribution of gluons in the proton, obtained from
studies of exclusive vector meson ($J/\psi, \rho$) production at HERA
(generalized parton distributions). Based on this information, we show
that the interaction of small--size color singlet configurations with
the proton approaches the ``black--disk limit'' (BDL) at small
$x$. This new dynamical regime is marginally visible in diffractive
DIS at HERA, and will be fully reached in central $pp$ collisions at
LHC, where it will have numerous qualitative implications for the
hadronic final state. In these proceedings we briefly summarize the
theoretical arguments leading to the BDL, and the evidence for it in
presently available small--$x$ data (HERA, Tevatron).  For details, as
well as for a discussion of the implications for $pp$ collisions at
LHC, we refer to Refs.~\cite{Frankfurt:2003td,Frankfurt:2005mc}.

\section{Dipole picture from QCD factorization}
At small $x$, $\gamma^\ast p$ scattering with longitudinal polarization
can be viewed as the scattering of small--size color--singlet quark/gluon 
configurations in the photon wave function from the proton.
Intuitively, this can be understood from the space--time evolution
in the target rest frame, in which the virtual photon typically fluctuates 
into a $q\bar q$ pair of transverse size $d \sim 1/Q$ a long time, 
$\tau_{\text{coh}} \sim 1/(2 m_N x)$, before hitting the target. 
Formally, it can be shown that the leading $\log Q^2$ approximation 
for the DIS cross section, $\sigma_L$, is equivalent to the scattering 
of a $q\bar q$ dipole from the proton 
with cross section \cite{Brodsky:1994kf,Frankfurt:1996ri} 
\begin{equation}
\sigma^{dp}
\;\; = \;\; 
\frac{\pi^2}{4} \; F^2 \; d^2 \; \alpha_s (Q_{\text{eff}}^2) \; 
x \, G(x, Q_{\text{eff}}^2) ,
\label{sigma_inel_LT}
\end{equation}
where $F^2 = 4/3$ (for $gg$ dipoles, $F^2 = 3$). The cross section,
which vanishes for small dipoles (``color transparency''), 
is proportional to the gluon density in the proton, 
evaluated at an effective scale $Q_{\text{eff}}^2 \approx (\pi / d)^2$.
We stress that, by the derivation from the factorized DIS cross section, 
the gluon density in Eq.~(\ref{sigma_inel_LT}) is unambiguously 
determined as the leading--twist gluon density in the leading 
$\log Q^2$ approximation. In particular, it is subject to DGLAP evolution,
which produces a strong rise at small $x$, and implies a rapid growth of 
the dipole cross section with increasing energy. The crucial question
is at which energies unitarity leads to a breakdown of this approximation.

Though restricted to the leading--log approximation, the dipole picture
with Eq.~(\ref{sigma_inel_LT}) is useful for discussing qualitative features 
of the HERA inclusive DIS data, such as the breakdown of the leading--twist 
approximation at small $x$ for $Q^2 \lesssim 4\, \text{GeV}^2$, and 
the difference between $\sigma_L$ and $\sigma_T$ \cite{Frankfurt:2005mc}.
The extension to diffractive final states requires one to explicitly 
consider $q\bar q g \ldots g$ configurations in the photon wave function.
Diffractive DIS thus allows one to probe the interaction of small--size 
quark/gluon configurations with hadronic matter in much more detail
than inclusive DIS.
\section{Transverse spatial distribution of gluons}
%
%
\begin{figure}[b]
\includegraphics[width=8cm]{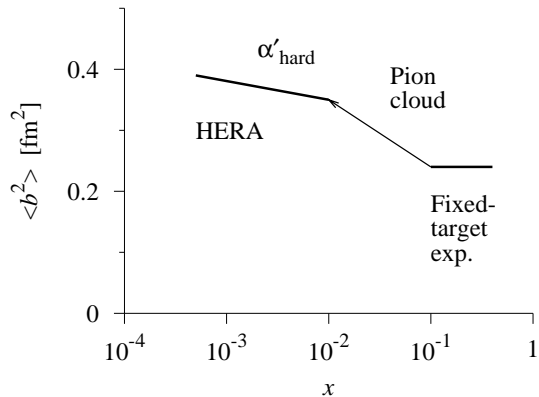}
\caption{Summary of the information on the proton's average 
transverse gluonic size, 
$\langle b^2 \rangle \equiv \int d^2 b \; b^2 \; F_g (x, Q^2; b)$,
from $J/\psi$ photoproduction ($Q^2 \approx 3\, \text{GeV}^2$). The
increase between $x \sim 10^{-1}$ and $10^{-2}$ can be attributed
to the pion cloud \cite{Strikman:2003gz}. For details see
Refs.~\cite{Frankfurt:2003td,Frankfurt:2005mc}.}
\label{fig:size}
\end{figure}
To study the role of unitarity in hard processes at small $x$,
we need to know not only the total density of gluons in the proton, 
but also their spatial distribution in the transverse plane.
This information comes from studies of exclusive vector meson production
($V = J/\psi, \rho$) in $\gamma^\ast_L p$ scattering at small $x$. 
On grounds of a general QCD factorization theorem, the amplitudes 
for these processes can be expressed in terms of the gluon generalized 
parton distribution (GPD) in the proton (here in the approximation
of zero ``skewness''),
\begin{equation}
G(x, Q^2; t) \;\; = \;\; G(x, Q^2) \; F_g (x, Q^2; t), 
\hspace{3em} F_g (x, Q^2; t = 0) \; = \; 1 . 
\end{equation}
$F_g$ is the normalized two--gluon formfactor, which can directly 
be extracted from the $t$--dependence of the differential cross section, 
$d\sigma (\gamma^\ast_L p \rightarrow Vp)/dt \propto F_g^2 (t)$.
Its Fourier transform with respect to the transverse momentum 
transfer to the proton, $\bm{\Delta}_\perp$,
\begin{equation}
F_g (x, Q^2; b) \;\; \equiv \;\; \int \frac{d^2 \Delta_\perp}{(2 \pi)^2}
\; e^{i (\bm{\Delta}_\perp \bm{b})}
\; F_g (x, Q^2; t)
\hspace{3em} (t = -\bm{\Delta}_\perp^2),
\label{rhoprof_def}
\end{equation}
describes the distribution of gluons (with longitudinal
momentum fraction $x$) with regard to transverse position, $\bm{b}$,
normalized according to $\int d^2 b \; F_g (x, Q^2; b) = 1$.

Extensive studies of $J/\psi$ photo/electroproduction at HERA
have demonstrated the applicability of the QCD factorization
formulae, with corrections due to the finite transverse size of the 
vector meson \cite{Frankfurt:2005mc}. Together with $J/\psi$ 
photoproduction data from fixed--target experiments
at lower energies, these data have produced a detailed picture of the
transverse spatial distribution of gluons at $Q^2
\approx 3 \, \text{GeV}^2$ over a wide range of $x$.
The proton's average transverse gluonic size increases with decreasing $x$, 
see Fig~\ref{fig:size}. Various dynamical mechanisms have been identified, 
which contribute to this growth in different regions of 
$x$ \cite{Frankfurt:2002ka,Strikman:2003gz}.
At higher $Q^2$, the growth with decreasing $x$ is slower, because
DGLAP evolution with increasing $Q^2$ effectively probes higher and higher
$x$ values in the input distribution at the starting scale.

An important observation is that the average transverse radius of the 
distribution of gluons with significant momentum fraction, 
$x \gtrsim 10^{-2}$, is \textit{considerably smaller} than the radius
of soft hadronic interactions in high--energy $pp$ collisions.
This implies that $pp$ events with hard processes involving two partons 
with $x_{1, 2} \gtrsim 10^{-2}$ (\textit{e.g.}, dijet production) 
originate from much more central collisions than minimum 
bias events \cite{Frankfurt:2003td}.

\section{Black--disk limit of dipole--proton scattering}
%
%
\begin{figure}[b]
\includegraphics[width=5cm]{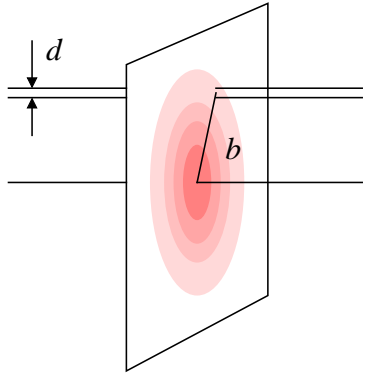}
\caption{Dipole--proton scattering in impact parameter representation}
\label{fig:dp_impact}
\end{figure}
On the basis of the dipole picture in the leading $\log Q^2$ approximation,
\textit{cf.}\ Eq.~(\ref{sigma_inel_LT}), and the information about the 
transverse spatial distribution of gluons, we can study the approach
to the unitarity regime in hard processes at small $x$. The problem
can be formulated in the language of an optical model for hadron--hadron 
scattering. We consider the elastic scattering amplitude for a 
small--size color dipole from the proton in impact
parameter representation (see Fig.~\ref{fig:dp_impact})
\begin{eqnarray}
A^{dp}(s, t) &=& \frac{i \, s}{4\pi} \int d^2 b \;
e^{-i (\bm{\Delta}_\perp \bm{b})}
\; \Gamma^{dp} (s, b) 
\hspace{3em} (t = -\bm{\Delta}_\perp^2),
\label{dp_impact}
\end{eqnarray}
where $\Gamma^{dp} (s, b)$ is the profile function. Using the optical 
theorem, the inelastic (total minus elastic) cross section can be
expressed in terms of the profile function as
\begin{equation}
\sigma_{\text{in}} (s)
\; = \; \int d^2 b \; \sigma_{\text{in}}(s, b),
\hspace{2em} \sigma_{\text{in}}(s, b) \; \equiv \;
1 - |1-\Gamma^{dp} (s,b)|^2 .
\label{sigma_inel}
\end{equation}
The function $\sigma_{\text{in}}(s, b)$ can be interpreted as the 
probability for inelastic interaction in dipole--proton scattering 
at a given impact parameter, $b$. It tends to unity if
\begin{equation}
\Gamma^{dp} \;\; \rightarrow \;\; 1.
\label{bdl}
\end{equation}
In optics, this limit corresponds to the scattering from 
a black disk, whence Eq.~(\ref{bdl}) is referred to as the 
``black--disk limit'' (BDL) of dipole--hadron scattering. 

\begin{figure}[t]
\includegraphics[width=10.5cm]{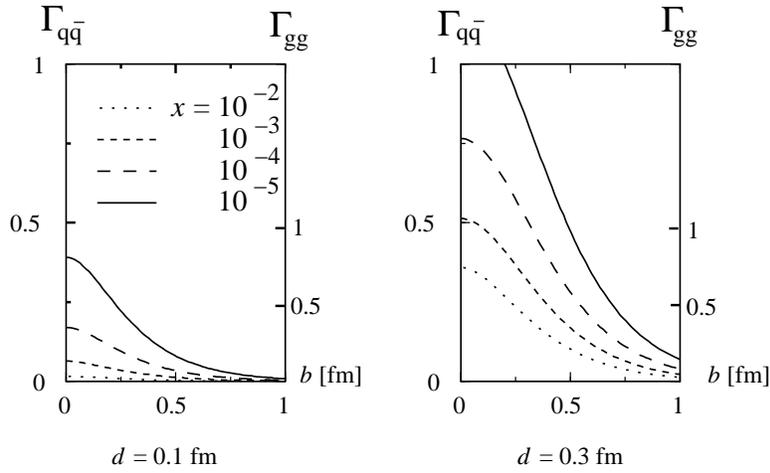}
\caption{The profile function for dipole--proton scattering, 
$\Gamma^{dp}(s, b)$, Eq.~(\ref{dp_impact}), 
as obtained from the LO DGLAP gluon density,
\textit{cf.}\ Eq.~(\ref{sigma_inel_LT}), and a phenomenological 
parametrization of the transverse spatial distribution of 
gluons \cite{Frankfurt:2003td}. Shown are the results for
dipoles of size $d = 0.1$ and $0.3\, \text{fm}$, and various
values of $x = \pi^2/(d^2 s)$. The left axes refer to 
$q\bar q$ dipoles, the right axes to $gg$ dipoles.}
\label{fig:profile}
\end{figure}

In the leading--twist (LT) approximation, the distribution 
$\sigma_{\text{in}}(s, b)$ is given by Eq.~(\ref{sigma_inel_LT}),
with the total gluon density replaced by the local density in
transverse space. While the LT formula is meaningful only 
as long as $\sigma_{\text{in}} (s, b) \ll 1$, we can use it to study the
\textit{breakdown} of the LT approximation and the onset 
of the BDL regime. Figure~\ref{fig:profile} shows the result for
$\Gamma^{dp}$ obtained in this approximation,
for both $q\bar q$ (left scale) and $gg$ (right scale) dipoles. 
One sees that the profile function approaches the BDL, Eq.~(\ref{bdl}),
at small impact parameters, even for small dipole sizes, once $x$ 
becomes sufficiently small. The reason is the growth of the gluon
density in the proton at small $x$ due to DGLAP evolution.

%
%

\section{Black--disk limit in $ep$ scattering at HERA}
Inclusive $\gamma^\ast_L p$ scattering in LO corresponds to the scattering
of a $q\bar q$ dipole from the proton. One sees that the BDL regime 
becomes marginally relevant at the upper end of the HERA energy range
($x \gtrsim 10^{-4}$) for $Q^2 \sim 4 \, \text{GeV}^2$ (corresponding to 
$d = 0.3 \, \text{fm}$). We emphasize that in the dipole picture 
only the longitudinal cross section, $\sigma_L$, 
is dominated by small--size dipoles. 
The transverse cross section, $\sigma_T$, receives significant 
contributions from large--size configurations in the virtual photon
even for substantial $Q^2$, and is thus not a suitable observable for 
probing the BDL in hard interactions. 

More sensitive to the onset of the BDL regime 
are diffractive processes in $\gamma^\ast p$ scattering,
which probe the interaction of $q\bar q g \ldots g$ dipoles
with the target. The fact that the probability 
for gluon--induced diffraction, $P_g$, is experimentally close
to its maximum value, 1/2, at the upper end of the HERA energy
range indicates that the interaction of such dipoles with the
proton is close to the BDL \cite{Frankfurt:2005mc}.

\section{Black--disk limit in $pp$ scattering at LHC}
Much smaller values of $x$ than in $ep$ scattering can be probed
in hard processes induced by leading partons in 
high--energy $pp$ collisions. At LHC ($\sqrt{s} = 14 \, \text{TeV}$),
a parton in the ``projectile'' proton with momentum fraction 
$x_1 \sim 10^{-1}$, in a hard collision producing a final state 
with transverse momentum $p_\perp = 2 \, \text{GeV}$, resolves
partons in the ``target'' with
\begin{equation}
x \;\; = \;\; 4 p_\perp^2 /(x_1 s) \;\; \sim \;\; 10^{-6} .
\end{equation}
In the dipole picture, such parton--parton processes can predominantly 
be associated with scattering of $gg$ dipoles from the target
(for which the cross section is $9/4$ times larger than for $q\bar q$
dipoles), with $Q^2 = (\pi / d)^2 = 4 \, p_\perp^2$. Under these 
conditions, the dipole--proton interaction for central collisions
is deep inside the BDL regime, see Figure~\ref{fig:profile}. 

The approach to the BDL qualitatively changes the interactions 
of leading partons in central $pp$ collisions. They
acquire substantial transverse momenta through their interaction
with the dense medium of small--$x$ gluons, 
$p_{\perp, \text{BDL}} \approx 4 - 5 \, \text{GeV}$ for $x_1 \sim 10^{-1}$
in the estimate of Ref.~\cite{Frankfurt:2003td}. As a result, one expects
significantly larger transverse momenta and lower multiplicity in forward
particle production, as well as other qualitative changes in the 
hadronic final state \cite{Frankfurt:2003td}. These effects can be
observed by selecting central $pp$ collisions with a trigger on hard dijet
production at central rapidities (see above) \cite{Frankfurt:2003td}.
Such measurements are feasible with the CMS and TOTEM detectors at LHC.

Evidence for the BDL in existing $pp/\bar p p$ data is the observation
that the $pp$ elastic scattering amplitude above the Tevatron energy 
($\sqrt{s} = 1.8 \, \text{TeV}$) seems to be ``black'' at central impact
parameters, $\Gamma^{pp}(b = 0) \approx 1$ \cite{Islam:2002au}. 
This is consistent with the fact that in the BDL all leading (valence) 
quarks acquire large transverse momenta due to hard interactions, 
making it impossible for the proton wave functions to 
remain coherent \cite{Frankfurt:2004fm}.
\\[.5ex]
{\bf Acknowledgments.} The research of M.~S.\ is supported by DOE.

\end{document}